\documentclass{ws-ijmpa}
\usepackage[super]{cite}
\usepackage{graphicx}
\begin{document}

\def\bq{\begin{eqnarray}}
\def\eq{\end{eqnarray}}


\markboth{N. Nicolaevici}
{Gauge dependence on QED amplitudes in expanding de Sitter space}

\catchline{}{}{}{}{}

\title{
Gauge dependence in QED amplitudes in expanding de Sitter space}

\author{\footnotesize Nistor Nicolaevici}

\address{Faculty of Physics, West University of Timi\c soara
\\
V. P\^arvan 4, 300223, Timi\c soara, Romania
\\
nicolaevici@email.ro}

\maketitle

\begin{history}
\received{Day Month Year}
\revised{Day Month Year}
\end{history}

\begin{abstract}

We consider first order transition amplitudes in external fields in QED in the expanding
de Sitter space and point out that they are gauge dependent quantities. We examine the gauge
variations of the amplitudes assuming a decoupling of the interaction at large times,
which allows to conclude that the source of the problem lies in the fact that the
frequencies of the modes in the infinite future become independent of the comoving momenta.
We show that a possibility to assure the gauge invariance of the external field amplitudes
is to restrict to potentials which vanish sufficiently fast at infinite times, and briefly
discuss a number of options in the face of the possible gauge invariance violation in the
full interacting theory.


\end{abstract}

\ccode{PACS Nos.: 04.65.+v}

\section*{1. Introduction}

de Sitter (dS) space has received a special attention in the study of quantum field
effects in curved spacetimes. It plays a fundamental role in inflationary cosmology models, and
due to its high symmetry it provides a convenient laboratory for obtaining closed form results.\,
dS space is also of special interest due to the fact that it allows a significant number of effects
which have no counterpart in the Minkowski space. For example,\, a well known result is that for the
minimally coupled free massless scalar field there is no dS-invariant vacuum state,\, and that in
certain vacua the quantum fluctuations can become arbitrarily large.\cite{vile,alle} Other\, notable
examples are that free particles can decay in multiple copies of\, themselves,
\cite{nach,dec1,dec2,dec3,maro} the interacting vacuum could be highly unstable and lead to a massive
particle production,\cite{myhr,ande5,poly1,poly2,krot,poly3,akhm5,akhm6} or the spacetime itself
could be instable due to quantum gravitational effects.\cite{ford,anto,tsam4,tsam5}

While the effects above have nothing intrinsically problematic, it also seems that dS space involves a series
of less trivial features, which indicate a departure from the flat space theory at a more fundamental
level: there are opinions that a well-defined $S$-matrix may not exist,\cite{witt,bous,boya}
infrared divergences in Feynman diagrams systematically show up,\cite{ford1,sasa,tsam7,seer,akhm1}
or unitarity might be violated by interacting fields.\cite{nonu,alva} The main intention of the present
paper is to point out another\, fact which seems to put the theory in dS space on an unequal footing with 
that in flat space.\, We will focus our attention on first order transition amplitudes in external fields
in spinor QED\, in the expanding patch of dS space. We will show that these amplitudes are gauge dependent
quantities. Despite the rather large amount of work on QED effects in dS space (see below), it appears 
that this fact has passed unnoticed. Here, we will be content to make clear that this is indeed the 
case, and briefly suggest a number of possibilities to deal with this situation.

Gauge invariance in QED calculations typically refers to the invariance of the $S$-matrix elements under
gauge transformations in (1) the wave functions of the external photon lines, (2) the external potential,
and (3) the photon propagator. Amplitudes at tree level for various\, QED processes in dS space involving
(1) and (2) in a particular gauge have been explicitly obtained in Refs.~\citen{tsar9,cruc0,cruc1,nico0,
cruc3,cota0,miha,cruc4,blag1,blag2}. Tree-level QED amplitudes in a flat FRW universe, which most probably
can be generalized to a dS space, can be also found in Refs.~\citen{birr7,lotz5,lotz6,lotz1, lotz2,lotz3,
lotz4,medv,tsar1,tsar2,tsar3,tsar4,nomu,higu1,kimu}.\, Loop calculations involving (3) with the photon
propagator in a dS invariant form in the Feynman gauge have been done in Ref.~\refcite{kahy1},\, in
the Landau gauge in Refs.~\citen{kahy2,prok2,prok3,prok1},\, and using a dS non-invariant analogue
of the Feynman gauge\cite{wood1} in\, Refs.~\citen{kahy1,kahy2,glav,leon1,leon2,wang1,wang2}. One
should stress that the evaluation of diagrams and implicitly the check of gauge invariance at higher
orders is generally significantly more
difficult than in flat space because of the lack of the machinery of Fourier transforms, and
also due to the more complex form of the propagators.\footnote{The photon propagator in a dS invariant
form in an arbitrary $R_\lambda$ gauge was only recently obtained in Ref.~\refcite{yous} The propagator  of
Allen and Jacobson\cite{alle1} corresponds\, to the\, Feynman gauge $\lambda =1$. The Landau gauge
$\lambda \rightarrow \infty$ is discussed in Ref.~\refcite{tsam}. See also
Refs.~\refcite{miao,frob1} for an arbitrary gauge.}

To our knowledge,\, the only systematic comparison between calculations in different gauges
in dS space was made in Ref.~\refcite{kahy2}, where the one-loop self-mass operator of a charged
scalar field was obtained both in the Landau\cite{tsam} and in the dS non-invariant
gauge.\cite{wood1} The self-masses in the two gauges differ, but this is nevertheless not enough
to conclude the violation of gauge invariance, which refers to physically measurable (on-shell)
quantities. Such a quantity can be obtained by considering the one-loop corrected evolution
of the scalar field.\cite{kahy2} It turns out that for an initial plane wave the evolution in the two gauges
leads to a qualitatively different time dependence at late times, which can be seen 
as a gauge dependence of the results. Another significant discrepancy was noted in Ref.~\refcite{kahy1},
where it was found that the self-mass in the Feynman gauge contains extra on-shell singularities
compared to that in the Landau gauge.

Our calculation will be considerably simpler, as we will focus on tree-level amplitudes. The plan
is as follows. We will consider transition amplitudes between the vacuum and a fermion-antifermion
state in an external potential $A_\mu$, with the particles in both the $in$ and $out$ states defined
by the Bunch-Davies modes. Similar amplitudes were used to discuss the possible unitarity
violation\cite{nonu, boya} and the vacuum decay\cite{higu, boya, alva} in the expanding dS space. In
particle production calculations, one usually introduces a different set of $out$ modes in order to
define the physical particles at late times, but as we will point out this is irrelevant to our question.
It will be then an easy task to show that for potentials of the form $A_\mu$=\,constant these
amplitudes do not vanish. Since such potentials correspond to a pure gauge, the
result is incompatible with the gauge invariance of the theory.\footnote{We have noted this
fact for the analogous amplitudes in scalar QED in a different context in Ref.~\refcite{nico}.}

We will make a further step and identify the mechanism which lies behind the gauge dependent amplitudes.
Introducing a decoupling of the interaction at infinite times, one can rewrite
the gauge variations of the amplitudes as an integral which contains the time derivative of the
decoupling factor. In the Minkowski space, the purely oscillatory time dependence of the modes
assures that this quantity always vanishes for an adiabatic decoupling. As we will see, this does not happen
in dS space. We will explain that the cause lies (not surprisingly) in the infinite expansion of space
in the infinite future, which makes all the frequencies of the modes approach the same value at
late times. The situation is somehow analogous to that of a transition with a vanishing Bohr
frequency, which can formally lead to divergent amplitudes. In our case, the pathology manifests
in the fact that the gauge variations of the amplitudes do not vanish for an adiabatic decoupling.

We will also propose a simple solution for assuring the gauge invariance of the external
field amplitudes. This essentially consists in restricting the calculations to external
potentials which vanish sufficiently fast at infinite times, in which conditions the gauge
variations of the amplitudes are rigorously zero. It remains to be seen with more calculations
in specific cases if our proposal is a satisfactory solution to the problem.

The paper is organized as follows. In Sec.~2 we establish the general form of the amplitudes.
In Sec.~3\, we exhibit the amplitudes which break the gauge invariance, and in Sec.~4 we identify
the mechanism which leads to such quantities. In Sec.~5  we discuss the condition for restoring
the gauge invariance of the amplitudes. We end in Sec.~6 by briefly listing a number of options
to deal with the possible gauge dependence in the fully quantized theory.

\section*{2. The transition amplitudes}

The line element of the expanding dS space is
\bq
ds^2 = dt^2 - e^{2Ht} d{\bf x}^2,
\label{dsmet1}
\eq
where $H>0$ is the expansion parameter. It is convenient to define the conformal time
\bq
\eta= -\frac{1}{H}\, e^{-Ht}, \quad \eta\in (-\infty, 0),
\eq
in terms of which the metric reads
\bq
ds^2=\frac{1}{(H\eta)^2}( d\eta^2 - d{\bf x}^2).
\label{metcon}
\eq

As is well known, in order to deal with spinor fields in curved spacetimes it is necessary to
introduce a local orthonormal frame $\{e_{\hat \alpha}\}$. We choose it to be the Cartesian
frame defined by (in conformal coordinates)
\bq
e_{\hat \alpha}^{\,\,\mu} =-(H\eta)\, \delta_\alpha^{\,\mu}.
\label{ortfra}
\eq
The first order QED amplitudes in the external potential $A_\mu$ have the general form
\bq
{\cal A}_{i\rightarrow f}=-ie \int d^4 x \sqrt {-g }\, \bar \psi_f\,
\gamma^{\alpha} \psi_i
A_{\hat \alpha},
\quad \,\,
A_{\hat \alpha} =e^{\,\,\mu}_{\hat \alpha}
A_{\mu},
\label{genamp}
\eq
where all notation is\, conventional. We will focus on the case when the initial and final wave
functions are solutions of the Dirac equation of a definite momentum ${\bf p}$ and helicity
$\lambda=\pm \frac {1}{2}$. It is useful to introduce
\bq
k = \frac{m}{H}, \quad
\nu_\pm= \frac{1}{2} \pm i\frac{m}{H},
\eq
where $m$ is the mass of the Dirac field. We also need the unit norm helicity two-spinors
$\xi_{\lambda}$ and $\eta_\lambda$
defined by
\bq
\quad
\frac{1}{2}\, ({\bf n}_p \cdot \mbox{\boldmath $\sigma$})\,\xi_\lambda({\bf p})
 =\lambda\, \xi_\lambda({\bf p}),
 \quad
\eta_\lambda({\bf p})=i\sigma_2\, \xi_\lambda^*({\bf p}),
\quad
{\bf n}_p=\frac{{\bf p}}{p}.
\eq
The solutions of the Dirac equation mentioned above can then be written
as\cite{mama, lyth, cota} ($\sigma \equiv 2\lambda)$:
\bq
u_{\bf p,\, \lambda}(\eta, {\bf x})
&=&
\frac{\sqrt{\pi p/H}}{2(2\pi)^{3/2}}
\times
(H\eta)^2
\left(
\begin{array}{c}
\,e^{+\frac{\pi k}{2}} H^{(1)}_{\nu_-}(-p\,\eta)\,\xi_\lambda({\bf p})
\\
\sigma
e^{-\frac{\pi k}{2}} H^{(1)}_{\nu_+}(-p\,\eta)\,\xi_\lambda({\bf p})
\end{array} \right)
e^{i{\bf p} {\bf x}}, \label{umod}
\eq
and
\bq
v_{\bf p,\, \lambda}(\eta, {\bf x})
&=&
\frac{\sqrt{\pi p/H}}{2(2\pi)^{3/2}}
\times
(H\eta)^2
\left(
\begin{array}{c}
-\sigma e^{-\frac{\pi k}{2}} H^{(2)}_{\nu_-}(-p\,\eta)\,\eta_\lambda({\bf p})
\\
e^{+\frac{\pi k}{2}} H^{(2)}_{\nu_+}(-p\,\eta)\,\eta_\lambda({\bf p})
\end{array} \right)
e^{-i{\bf p} {\bf x}}, \label{vmod}
\eq
where $H^{(1,2)}_\nu(z)$ are the Hankel functions of the first and second kind. The four-spinors
are in the standard Dirac representation with the matrix $\gamma^0$ diagonal. The solutions (\ref{umod})
and (\ref{vmod}) can be identified as positive and negative frequency modes, respectively, in the
sense that in the infinite past they are purely oscillatory in the conformal time,
i.e.\footnote{The vacuum defined by these modes is the Bunch-Davies vacuum\cite{birr}.}
\bq
u\sim e^{-ip \eta}, \quad
v\sim e^{ip \eta}, \quad \eta\rightarrow -\infty.
\eq

It is convenient for our discussion to consider amplitudes for particle production from
the initial vacuum. The wave functions $\psi_f$ and $\psi_i$ have then to be identified
with $u_{{\bf p},\, \lambda}$ and $v_{{\bf p}^\prime,
\, \lambda^\prime}$, in which conditions
\bq
{\cal A}({\bf p}, {\bf p}^\prime)_{\lambda \,\lambda^\prime}=-ie \int d^4 x \sqrt {-g }
\,
\bar u_{{\bf p},\, \lambda}
\gamma^{\alpha}
v_{{\bf p}^\prime,\, {\lambda^\prime}}\,
A_{\hat \alpha}.
\label{auvgen}
\eq
It will also be sufficient to restrict to potentials $A_\mu$ of the following form:
\bq
\quad A_0=0,\quad  A_i=A_i(\eta),
\quad i=1,2,3.
\label{agen}
\eq

Let us write Eq. (\ref{agen}) in a more explicit way. Notice that thanks to the factorizable
dependence on $\eta$ and ${\bf x}$ in Eqs. (\ref{metcon}), (\ref{ortfra}), (\ref{umod}), (\ref{vmod})
and (\ref{agen}) the temporal and spatial integrations can be separately performed. The integration
with respect to ${\bf x}$ is immediate and leads to a delta-Dirac function. A simple calculation
then shows that the remaining factors can be organized as follows ($\sigma^i$ are the Pauli
matrices):
\bq
{\cal A}({\bf p}, {\bf p}^\prime)_{\lambda \,\lambda^\prime}=-ie \delta^3
({\bf p}+{\bf p}^\prime) [\,\xi^+_\lambda({\bf p})
\sigma^i\,
\eta_{\lambda^\prime}({\bf p}^\prime)] \, F_i (p)_{\sigma \sigma^\prime},
\label{AFgen}
\eq
where we collected the integrals with respect to $\eta$ in
\bq
&F_i(p)_{\sigma \,\sigma^\prime}&=f_i^+(p)-\sigma \sigma^\prime f_i^-(p),
\label{defF}
\\
&f_i^\pm(p)&=\frac{\pi p}{4}\, e^{\pm k\pi}
\int_{-\infty}^0 d\eta\, \eta A_i(\eta) [ H_{\nu_\pm}^{(2)}(-p\,\eta)]^2.
\label{fint}
\eq
It is clear that the specific `dS part' of the amplitudes is encoded in the $\eta$-integrals
(\ref{fint}). One can easily check that in the flat space limit $H\rightarrow 0$ the amplitudes
(\ref{fint}) reduce, as expected, to their analogues
in the Minkowski space.\footnote{The limits for the Hankel functions with $\nu_\pm=\frac{1}{2}\pm ik$
can be found e.g. in Eq. (5.6) of Ref.~\refcite{nach}.}

\section*{3. Gauge dependence of the amplitudes}

We now show that the amplitudes (\ref{AFgen}) are not gauge invariant. Let us choose
an external potential of the form
\bq
\quad A_0=0, \quad A_i=\mbox{constant}.
\label{purgau}
\eq
The corresponding field strength $F_{\mu\nu}=\partial_\mu A_\nu-\partial_\nu A_\mu$ is
identically zero, so that the potential is\, a pure gauge. Gauge invariance in these conditions
requires the amplitudes to vanish.

Let us look more closely at Eq. (\ref{AFgen}) in this case. Since the directions of
${\bf p}$ and $A_i$ are arbitrary, the vanishing of ${\cal A}({\bf p}, {\bf p}^\prime)_{\lambda
\,\lambda^\prime}$ for all helicities $\lambda$, $\lambda^\prime$ requires $F_i ({\bf p})_{\sigma \,
\sigma^\prime}$ to vanish. The arbitrariness of $\sigma$, $\sigma^\prime$ further implies that
$f_i^\pm({\bf p})$ must vanish. For the potentials (\ref{purgau}) these functions are
\bq
f_i^\pm( p)
\sim
\int_{-\infty}^0 d\eta\, \eta [ H_{\nu_\pm}^{(2)}(-p\,\eta)]^2.
\label{fgauint}
\eq
It is clear that these integrals are not identically zero. This means that the amplitudes
cannot be gauge invariant.

For safety, let us make sure that the integrals (\ref{fgauint}) are not ill-defined. The behavior of
the functions $H_\nu^{(2)}(z)$ for small and large arguments $z$
is
\bq
H_\nu^{(2)}(z)\simeq -\frac{i}{\pi}\, \Gamma (\nu)
\left ( \frac{z}{2}\right )^{-\nu}\!\!\!\!,
\,\,
z \rightarrow 0,
\quad
H_\nu^{(2)}(z)\simeq \sqrt{\frac{2}{\pi z}}\, e^{-i(z-\frac{\pi}{2}\, \nu - \frac{\pi}{4})},
\,\,
z \rightarrow \infty.
\label{hankz}
\eq
The first relation shows that for $\eta \rightarrow 0$\, the integrand is $\sim \eta^{\pm 2ik}$,
which puts no convergence problems. The second one implies that for $\eta \rightarrow -\infty$\,
the integrand is a pure phase $\sim e^{2ip\,\eta}$, which is also not problematic
(the integral can be made well-defined by introducing a small convergence factor,
corresponding to a decoupling of the interaction in the far past). For example, a
simple way to see that $f_i^\pm(p)$ are not identically zero is by
considering $p\rightarrow \infty$. Using the second relation in Eq. (\ref{hankz})
the integration is immediate and one finds $f_i^\pm(p)\sim p^{-1}$.

It is worth recalling how gauge invariance is ensured in the Minkowski space. Considering
as before a potential $A_\mu$=\,constant, the time dependence under the integral analogous to
Eq. (\ref{auvgen}) is completely determined by the purely oscillatory factors
$\bar u_{{\bf p},\, \lambda}\sim e^{iE_{{\bf p}}\, t},\quad v_{{\bf p}^\prime,\,
\lambda^\prime}\sim e^{iE_{{\bf p}^\prime}\, t}$, which implies
\bq
{\cal A}({\bf p}, {\bf p}^\prime)_{\lambda \,\lambda^\prime}\sim
\delta (E_{{\bf p}}+E_{{\bf p}^\prime}),
\eq
so that the amplitudes identically vanish. This is rather trivial, so let us recall the
case of a  general gauge transformation
\bq
A_\mu \rightarrow A^\prime_\mu=A_\mu+ \partial_\mu \Lambda.
\label{gengau}
\eq
In these conditions the gauge variations of the amplitudes are
\bq
\Delta {\cal A}({\bf p}, {\bf p}^\prime)_{\lambda \,\lambda^\prime}
=-ie \int d^4 x \,
\bar u_{{\bf p},\, \lambda}\gamma^{\mu}v_{{\bf p}^\prime,\, {\lambda^\prime}}\,
(\partial_\mu \Lambda).
\label{genvar}
\eq
Assuming that $\Lambda(x)$ can be Fourier transformed,
\bq
\Lambda(x)=\int d^4 k\, \hat \Lambda (k)\, e^{ikx},
\eq
the Fourier components of Eq. (\ref{genvar}) are
\bq
\sim \delta^4(p+p^\prime -k)\bar u_{\lambda}({\bf p})
(k_\mu\gamma^{\mu})\, v_{\lambda^\prime}({\bf p}^\prime),
\label{foucon}
\eq
which again vanish due to the on-shell relations
\bq
(p_\mu\gamma^{\mu}
-m) u_\lambda({\bf p})=0,
\quad
(p_\mu\gamma^{\mu} +m) v_\lambda ({\bf p}) =0.
\label{shell}
\eq
This calculation might leave the impression that gauge invariance in dS space is lost due to
the absence of on-shell relations like Eqs. (\ref{shell}), as implied by the time-dependent
metric. This is however not so: the vanishing of Eq. (\ref{foucon}) is actually the Fourier
transform of the current-conservation-like relation
\bq
\partial_\mu(\bar u \gamma^\mu v)=0,
\label{cons}
\eq
which admits the curved spacetime generalization, i.e.
\bq
\partial_\mu(\sqrt {-g }\, e^{\,\,\mu}_{\hat \alpha}
\bar u \gamma^\alpha v)=0.
\label{consds}
\eq
In the next section we will reexamine the problem based on Eq. (\ref{consds}), which will allow
to precisely identify what goes wrong in dS space.

\section*{4. The adiabatic residue}

We first recall some known facts from the Minkowski space. A usual way to check the invariance of
the amplitudes (\ref{auvgen}) under the gauge transformations (\ref{gengau}) in flat space is to
perform an integration by parts and then use Eq. (\ref{cons}). However, the desired property
rigorously follows
only if one can ignore the surface terms.\footnote{The potential $A_\mu$ is not a physical field, so
there is no reason to assume that $\Lambda$ generally vanishes at the boundary of spacetime.} The
contributions from spatial infinity can be ignored with the usual assumption that at infinite distances
the physical fields vanish, but this is not allowed for the hypersurfaces at $t\rightarrow \pm \infty$.
One way to deal with the problem is to decouple the interaction at infinite times, which allows to
ignore the surface terms. After this step, the gauge variations of the amplitudes are given by an
integral which contains the time derivative of the decoupling factor. In the Minkowski space, the key fact
is that for an adiabatic decoupling the purely oscillatory form of the modes always eliminates this
term. Let us see how this works in dS space.

We denote the decoupling functions by $h_\epsilon(t)$, where $\epsilon$\, is the decoupling
parameter. We request as usual
\bq
\lim_{\epsilon\rightarrow 0} h_{\epsilon}(t)=1
\,\,\,\mbox{for $t$ fixed},
\quad \quad
\lim_{t\rightarrow \pm \infty} h_{\epsilon}(t)=0
\,\,\,\mbox{for $\epsilon>0$ fixed}.
\label{decgen}
\eq
For clarity, we begin with the Minkowski case. Introducing the decoupling function in
Eq. (\ref{genvar}) and performing the integration by parts one finds
\bq
\Delta {\cal A}({\bf p}, {\bf p}^\prime)_{\lambda \,\lambda^\prime}
=ie \int d^4 x \,
h^{\,\prime}_\epsilon(t)\,
\bar u_{{\bf p},\, \lambda}
\gamma^{0}
v_{{\bf p}^\prime,\, {\lambda^\prime}}\,
\Lambda,
\label{varmin}
\eq
where the prime denotes derivation with respect to the argument. Gauge invariance requires
that in the adiabatic limit $\epsilon \rightarrow 0$\, this quantity must vanish. We make
some simplifications at this step. It is evident that the vanishing of Eq. (\ref{varmin})
can only result from the integral with respect to $t$, i.e.
\bq
\sim \int_{-\infty}^\infty dt\, h^\prime_\epsilon(t)\,
e^{i (E_{\bf p}+iE_{{\bf p}^\prime}) t} \Lambda (t).
\label{basint}
\eq
It is also clear that the vanishing property has nothing to do with the form of $\Lambda(t)$,
so let us assume that $\Lambda(t)$ is\, zero for $t<0$ and $\Lambda(t) =1$ for $t\geq 0$.
In these conditions the integral reads
\bq
R_M(\epsilon) \equiv
\int_0^{\infty} dt\, h^\prime_\epsilon(t)\,
e^{i (E_{\bf p}+E_{{\bf p}^\prime}) t}.
\label{reps}
\eq
For example, for an exponential decoupling
\bq
h_\epsilon(t)=e^{-\epsilon \vert t\vert},
\quad
R_M(\epsilon)=\frac{-i\epsilon}{E_{\bf p}+E_{{\bf p}^\prime} -i\epsilon},
\label{expdec}
\eq
and as expected the result vanishes for $\epsilon \rightarrow 0$. The same test can be applied
to any spacetime. We will call integrals similar to Eq. (\ref{reps}) in the
limit $\epsilon \rightarrow 0$\, $adiabatic$ $residues$. Gauge invariance of the amplitudes
can then be translated by saying that
the adiabatic residues must vanish.

We now apply the test to dS space. The first step is to establish the analogue of Eq. (\ref{reps}).
We introduce in the general form of the amplitudes (\ref{genamp}) the decoupling functions $h_\epsilon(t)$
and consider the gauge variations implied by the transformations (\ref{gengau}).
Integrating by parts using Eq. (\ref{consds}) one finds
\bq
\Delta {\cal A}({\bf p}, {\bf p}^\prime)_{\lambda \,\lambda^\prime}
=ie \int d^4 x \sqrt {-g }\,\, h^\prime_\epsilon(t)\, e_{\hat \alpha}^{\,\,0}\,
\bar u_{{\bf p},\, \lambda} \gamma^{\alpha} v_{{\bf p}^\prime,\, {\lambda^\prime}}\, \Lambda.
\label{vards}
\eq
We are interested in the integral with respect to $t$ in the limit $\varepsilon \rightarrow 0$. Note
that, in contrast to the Minkowski case, there is now an essential past-future asymmetry, as implied
by the expansion of space. The important observation is that in the adiabatic limit
the derivative of the decoupling  function is $h_\epsilon^\prime(t)\rightarrow 0$ (for $t$ fixed),
which implies that the integral can be non-zero only due to the contributions from infinite times
$t\rightarrow \pm \infty$. Furthermore, since for $t\rightarrow -\infty$ the
modes become purely oscillatory with respect to $\eta$, the picture in this limit is essentially
the same with that in the Minkowski space. As a consequence, a non-zero adiabatic residue can
only come from the contributions from $t\rightarrow +\infty$. This means that we can make
the same choice for $\Lambda(t)$ as in the calculation above.

From now on we consider all quantities as  functions of the time $\eta$. Notice that
the relevant integration interval $t\in [0, \infty)$ translates into
\bq
\eta\in [-H^{-1}, 0).
\label{varcon}
\eq
Using the explicit form of the modes (\ref{umod}) and (\ref{vmod}) one finds that the
integral with respect to $\eta$ in Eq. (\ref{vards}) is a linear combination of\, integrals of the
following form:
\bq
\sim \int_{-H^{-1}}^0 d\eta\, h^\prime_\epsilon(\eta)\eta\, H^{(2)}_{\nu_\pm}(-p\,\eta)
H^{(2)}_{\nu_\mp}(-p^\prime\eta).
\label{casint}
\eq
The distinction between the $\pm$ cases is inessential, as it amounts to $p\leftrightarrow p^\prime$.
Hence, the analogue of Eq. (\ref{reps}) can be chosen to be
\bq
R_{dS}(\epsilon)\equiv p\int_{-H^{-1}}^0
d\eta\, h^\prime_\epsilon(\eta)\eta\, H^{(2)}_{\nu_+}(-p\,\eta) H^{(2)}_{\nu_-}(-p\eta),
\label{rds}
\eq
where for simplicity we set $p^\prime=p$. The factor in front of the integral was introduced only
for making the expression dimensionless. One can check that in the flat space limit $H\rightarrow 0$
the integral expressed in terms of the time $t$ reduces to the Minkowski form (\ref{reps})
with $E_{\bf p^\prime}=E_{\bf p}$ (up to an inessential factor depending on $p$).

It is interesting to explicitly obtain the adiabatic residue implied by Eq. (\ref{rds}) for the
exponential decoupling (\ref{expdec}). The functions $h_\epsilon$
in terms of the time $\eta$ in the interval of interest are
\bq
\qquad
h_\epsilon(\eta)=(-H \eta)^{\,\frac{\epsilon}{H}},
\quad
-H\eta \in (0, 1),
\label{deceta}
\eq
in which conditions
\bq
R_{dS}(\epsilon)= \frac{p}{H} \,
\epsilon
\int_{-H^{-1}}^0 d\eta (-H \eta)^{\,\epsilon/H}
H^{(2)}_{\nu_+}(-p\,\eta) H^{(2)}_{\nu_-}(-p\eta).
\label{HH}
\eq
The integral (\ref{HH}) can be evaluated in a straightforward way by expanding the product of Hankel
functions as a power series in $\eta$ using ($z\equiv -p\,\eta$)
\bq
H_\nu^{(2)}(z)=\frac{e^{\nu \pi i} J_\nu(z)-J_{-\nu}(z)}{i \sin \pi \nu},
\quad
J_\nu(z)=
\left(\frac{z}{2}\right)^\nu
\sum_{n=0}^\infty \frac{(-1)^n}{n!\, \Gamma(\nu+n+1)}
\left( \frac{z}{2}\right)^{2n}.
\eq
The result  is rather complicated, but it significantly simplifies for $\epsilon \rightarrow 0$.
We first observe that the power expansion is of the form
\bq
\qquad\quad\quad  H^{(2)}_{\nu_+}(z) H^{(2)}_{\nu_-}(z)=\frac{A_H}{z}\, +\, \mbox{Rest}(z),
\quad A_H=
-\frac{2}{\pi}\frac{1}{ \cosh \left(\frac{\pi m}{H}\right)},
\label{dive}
\eq
\bq
\mbox{Rest}(z) = z^{2ik} B(z) +z^{-2ik} C(z) + z D(z),\qquad\,\,\,\,\,
\label{rest}
\eq
where the functions $B$, $C$ and $D$ are sums over $non$-negative integer powers of $z$.
Another essential observation is that the factor $\sim\epsilon$ in front of the integral (\ref{HH})
implies that the limit $\epsilon \rightarrow 0$ is determined only by the terms
in the series which lead to a divergent quantity $\sim \epsilon^{-1}$. It is easy to see that
such a quantity can be produced only by the term $\sim \frac{1}{z}$ in Eq. (\ref{dive}). This
leaves us with
\bq
\lim_{\epsilon \rightarrow 0}R_{dS}(\epsilon)
=A_H \lim_{\epsilon \rightarrow 0} \,\epsilon \int_{-H^{-1}}^0 d\eta (-H \eta)^{\,\epsilon/H-1}
\qquad\qquad\,
\nonumber
\\
=
-A_H \lim_{\epsilon \rightarrow 0}\, (-\eta H)^{\,\epsilon/H}\,
{\Big \vert}_{\eta=-H^{-1}}^{\eta=0}=
A_H.\,\,\,\,\,\,\quad
\label{ards}
\eq
We have thus obtained that the adiabatic residue in dS space does not vanish. This shows
that the dS amplitudes cannot be gauge independent quantities. Note that, as expected, in
the limit $H\rightarrow 0$ we recover the vanishing result from the flat space. A
representation of $R_{dS}(\epsilon)$ defined by Eq. (\ref{HH}) as a function of\, $\epsilon$
for different values of $\frac {H}{m}$\, is shown in Fig. 1.

\begin{figure}

\centerline{\includegraphics[width=3.8in, height=5.2in, angle=90]{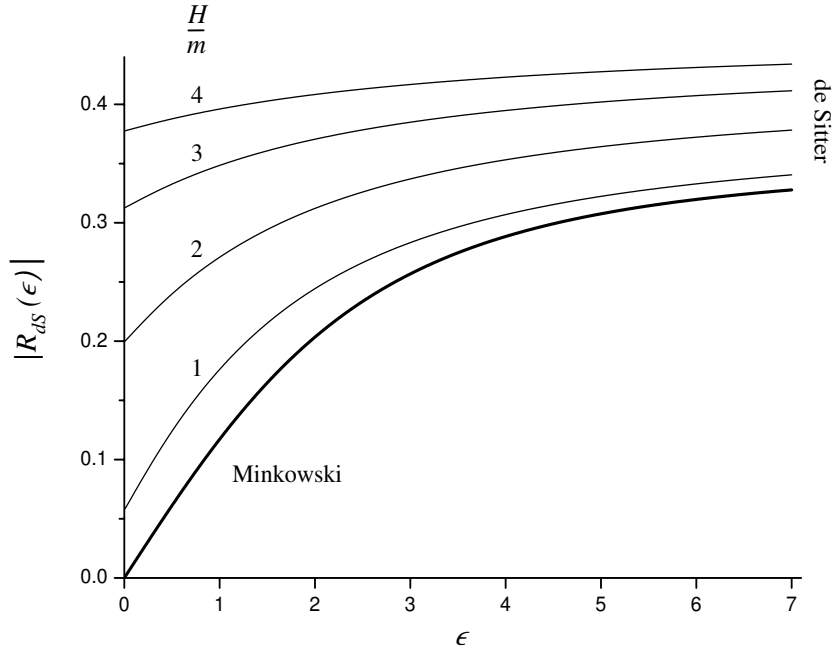}}

\caption{The absolute value of Eq. (\ref{HH}) represented as a function of $\epsilon$
for different ratios $\frac{H}{m}$ (shown near the curves) and $\frac{p}{m}=1$. The curve
in bold corresponds to the flat space limit $H\rightarrow 0$.}

\end{figure}

It is remarkable that the same result (\ref{ards}) is obtained irrespective of the form of the decoupling
functions. In order to see this it is sufficient to go back to Eq. (\ref{rds}) and repeat the same steps
using Eqs. (\ref{dive}) and (\ref{rest}) for an arbitrary $h_\epsilon(\eta)$. One similarly finds that
for $\epsilon \rightarrow 0$ the integral reduces to
\bq
\lim_{\epsilon \rightarrow 0}
R_{dS}(\epsilon)=-A_H \lim_{\epsilon \rightarrow 0}\,
\int_{-H^{-1}}^0 d\eta\, h^\prime_\epsilon(\eta)=
-A_H \lim_{\epsilon \rightarrow 0}\, h_\epsilon(\eta){\Big \vert}_{-H^{-1}}^0=A_H.
\label{newred}
\eq
The last identity follows from
\bq
\lim_{\epsilon \rightarrow 0}\,h_{\epsilon}(\eta<0)=1, \quad
\lim_{\eta\rightarrow 0} h_\epsilon(\eta)=0\, \vert_{\epsilon>0},
\eq
which are the analogue of Eq. (\ref{decgen}) in terms of the time $\eta$.

Let\, us rephrase our calculation. In Minkowski space the vanishing of the adiabatic residue
is assured by the fact that the oscillatory form of the modes with respect to time keeps the
integral (\ref{basint}) finite, which in the adiabatic limit forces the integral to vanish.
This does not happen in dS space, and from the calculation above it
is clear that the cause lies in the behavior of the modes for $\eta\rightarrow 0$, or equivalently
infinite cosmological times
$t\rightarrow \infty$. It is useful to look at the contributions from this limit in
$R_{dS}(\epsilon)$ in the following way. Note that a purely oscillatory behavior with respect
to $t$ appears in terms of $\eta$ as
\bq
e^{\mp i\omega t} \sim \eta^{\pm \frac{i\omega}{H}}.
\eq
The non-zero result (\ref{newred}) is produced only by the divergent non-oscillatory term
$\sim \frac{1}{\eta}$ in Eq. (\ref{dive}). The rest of the terms (\ref{rest}) are finite oscillatory
terms, which are irrelevant in the adiabatic limit. It is crucial to observe that if one replaces
$\frac{1}{\eta}$ in Eq. (\ref{dive}) with a divergent oscillatory term $\sim \eta^{-1+i\alpha}$
(with $\alpha$ real) the integral (\ref{HH}) remains finite for $\epsilon\rightarrow 0$,
which implies a vanishing adiabatic residue. Hence, in a technical sense, gauge dependence in dS space
arises due to the fact that in the integral which defines the gauge variations of the amplitudes
the integrand looses its oscillatory behavior at
$\eta\rightarrow 0$.

We considered in our discussion that the initial and final wave functions in the amplitudes
are of the form $\bar \psi_f\psi_i=\bar u v$. It turns out that the situation is practically
the same for\, the combinations\, $\bar u u$ and $\bar vv$ (corresponding to scattering amplitudes).
The only difference in these cases is that the product of Hankel functions in Eq. (\ref{rds})
is replaced by $H^{(1)}_{\nu_\pm}(-p\,\eta) H^{(2)}_{\nu_\mp}(-p^\prime\eta)$. An identical
calculation shows that the expansion in powers of $z$ has the same form (\ref{dive}), so that
the adiabatic residue is non-zero in these cases too. It is interesting that for all possible
$u$-$v$ combinations the divergent term which is responsible for the non-zero adiabatic residue
is  a quantity of the form
($a,b=1$ or 2)
\bq
\quad
H^{(a)}_{\nu_\pm}(-p\,\eta) H^{(b)}_{\nu_\mp}(-p^\prime\eta)
\sim \eta^{-(\nu_++\nu_-)}=\eta^{-1}, \quad \eta\rightarrow 0.
\label{univ}
\eq
This `universal' dependence on $\eta$ will be essential when establishing the condition
for ensuring the gauge invariance of the amplitudes in Sec.~5.

Let us stress that the oscillatory behavior of the integrand for $\eta\rightarrow \infty$
in Eq. (\ref{casint}) does not disappear due to the modes themselves, which remain oscillatory
in the infinite future. One can easily check that in this limit both types of modes
(\ref{umod}) and (\ref{vmod})  contain components which oscillate as
\bq
\qquad
\eta^{\pm ik}\sim e^{\mp i m t}, \quad t\rightarrow \infty.
\label{latfre}\eq
The key fact is that the limit behavior (\ref{latfre}) does not depend on the momentum of the modes, which
is what leads to the disappearance of the oscillatory behavior for $all$ pairs of initial and final
momenta ${\bf p}$, ${\bf p}^\prime$. The fact that the late time frequencies become independent of the
comoving momenta can be obviously recognized to be an effect of the arbitrarily large expansion of space
at $t \rightarrow \infty$. Thus, from a physical point of view, the loss of gauge invariance of the
amplitudes is a consequence of this property.

An immediate generalization of the conclusion above is that an identical situation can be expected to
occur in a FRW spacetime with the same behavior at late times. It is also evident
that if the physical process of interest is sufficiently localized in time so that the arbitrarily
large expansion of space at large times can be ignored (the case of usual experiments) the problem
will not appear.

For completeness, in Appendix A we included the analogous calculation for the
adiabatic residue in scalar QED. The result is very similar to that in Eq. (\ref{ards}).

\section*{5. Restoring the gauge invariance}

We have remarked in a previous work \cite{nico} in the context of scalar QED calculation in the
same background that the gauge invariance of the amplitudes analogous to Eq. (\ref{auvgen}) is
assured if one restricts to potentials $A_\mu$ which vanish in the infinite future, i.e.
\bq
\lim_{\eta \rightarrow 0} A_\mu(\eta, {\bf x})=0.
\label{conpot}
\eq
Let us show that the same prescription can be applied for spinor QED.

The proof is practically the same with that in Minkowski space, and consists in showing that for
the potentials (\ref{conpot}) the problematic surface term from $\eta\rightarrow 0$ which was neglected
in the integration by parts (\ref{vards}) can be ignored without introducing the decoupling
functions. Considering the general form of the amplitude (\ref{genamp}), the surface
term of interest is (we continue to use the conformal coordinates)
\bq
\Delta {\cal A}
_{i\rightarrow f}
=-ie
\lim_{\eta\rightarrow 0} \int d^3{\bf x}
\sqrt {-g }\, e_{\hat \alpha}^{\,\,0}\,
\bar
\psi_f \gamma^{\alpha}
\psi_i\, \Lambda.
\label{boulim}
\eq
The only difference with respect to the Minkowski case is that we must be careful about the
possible divergences on the integration hypersurface. One finds that for any
combination of $u$-$v$ modes in the initial and final states the integrand for
$\eta\rightarrow 0$ behaves as (compare with Eq. (\ref{univ})):
\bq
\qquad \sqrt {-g }\, e_{\hat\alpha}^{\,\,0}\,
\bar \psi_f \gamma^{\alpha} \psi_i\, \sim \,
\eta H^{(a)}_{\nu_\pm}(-p\,\eta) H^{(b)}_{\nu_\mp}(-p^\prime\eta)
\sim \eta^{1- (\nu_++\nu_-)}=1.
\label{limint}
\eq
The important\, fact from Eq. (\ref{limint}) is that for the vanishing of the surface term
(\ref{boulim}) it is sufficient for the function $\Lambda$ to vanish at $\eta=0$. We observe
at this point that the gauge transformations are actually not determined by $\Lambda$, but
only by the derivatives
\bq
\Lambda_\mu \equiv \partial_\mu \Lambda.
\eq
By restricting to the potentials (\ref{conpot}) the same condition must be respected by $\Lambda_\mu$,
and in these conditions one can always consider that $\Lambda$ has the property mentioned above. This
can be easily seen with the redefinition
\bq
\Lambda(\eta, {\bf x})\rightarrow
\Lambda^\prime(\eta, {\bf x})=\int_{(0,\, {\bf x}_0)}^{(\eta,\, {\bf x})}
d x^\mu \Lambda_\mu,
\eq
where $(0, {\bf x}_0)$ is a fixed point on the hypersurface $\eta=0$, and where the line integral runs
along an arbitrary curve which connects the two points ($\partial_\mu\Lambda_\nu=\partial_\nu\Lambda_\mu$).
By construction, the new partial derivatives are $\Lambda^\prime_\mu=\Lambda_\mu$, while on the integration
hypersurface $\Lambda^\prime (0, {\bf x})=0$, which ends the proof.

An undesirable feature of Eq. (\ref{conpot}) is that it is not a gauge invariant relation.
This can be remedied if one strengthens the condition by requesting $A_\mu$ to smoothly vanish with
respect to $\eta$, i.e.
\bq
\lim_{\eta \rightarrow 0} \partial_\eta A_\mu(\eta, {\bf x})=0.
\label{strcon}
\eq
Equations (\ref{conpot}) and (\ref{strcon}) combined imply $\lim_{\eta \rightarrow 0}
\partial_\mu A_\nu(\eta, {\bf x})$, from which
\bq
\lim_{\eta \rightarrow 0}F_{\mu\nu}(\eta, {\bf x})=0,
\label{confie}
\eq
which is in a gauge invariant form. One can easily show that the last condition is essentially
equivalent to the first two ones, in the sense that if the electromagnetic tensor respects
Eq. (\ref{confie}) one can always choose a potential which respects Eqs. (\ref{conpot}) and
(\ref{strcon}).\cite{nico}

We emphasize that all the conditions above are expressed in terms of the conformal coordinates
$(\eta, {\bf x})$. The limits will generally assume a different form in other coordinate systems.
For example, if one replaces $\eta$ with the time $t$ the temporal
component in Eq. (\ref{conpot}) becomes (the other components remain unchanged)
\bq
\lim_{t \rightarrow \infty}
\left(
e^{Ht} A_t(t, {\bf x})\right)=0,
\eq
which implies a much faster vanishing of $A_t$ with respect to $t$.

We agree that these conditions might\, appear too restrictive, especially having in mind
that in flat space one can deal with even divergent potentials at infinite times. One can
take the view that
they are the price to be paid for the rather extreme conditions implied by the infinite expansion
of space at $t\rightarrow \infty$. It remains to be seen with more concrete calculations if
our proposal is indeed a solution to the problem.

\section*{6. Conclusions and further proposals}

There exists a lot of work on QED effects in dS space. However, it seems that the question of the
gauge invariance has not yet received a detailed investigation. In this paper we brought evidence
that in the expanding patch of dS space gauge invariance does not necessarily holds. Our conclusion
comes from considering tree-level amplitudes in an external field, with the initial and
final particle states defined by the Bunch-Davies modes. A simple illustration is provided by
the amplitudes for pure gauges of the form $A_\mu$=constant. One finds that these amplitudes do
not generally vanish, which is incompatible with the gauge invariance of the theory.

For a more general analysis, we examined the gauge variations of the amplitudes assuming a decoupling of
the interaction at infinite times. The gauge variations are then contained in an integral
which contains the time dependent factors of the modes and the time derivative of the
decoupling function. In the Minkowski space this integral always vanishes for an adiabatic
decoupling, which is assured by the purely oscillatory time dependence of the modes. This does
not happen in the expanding dS space, and the cause lies in the fact that the integrand becomes
non-oscillatory in the infinite future. However, this does not result from the fact that the modes
cease to oscillate in this limit, but due to the special dependence on the oscillatory
components in the Dirac current $\bar \psi\gamma^\mu\psi$, combined with the fact that the late time
oscillatory behavior becomes independent on the comoving momenta. Hence, the loss of gauge invariance
in dS space can be identified to be an effect of the infinite expansion of space at $t\rightarrow \infty$.
This suggests that the same problem will appear in more general FRW spaces with the same behavior
at large times. The problem, however, will not appear if for the physics of interest
the arbitrarily large expansion of space at late times can be ignored.

We suggested a possible solution for assuring the gauge invariance of the amplitudes, which
consists in restricting the calculations to potentials which in conformal coordinates respect
$\lim_{\eta \rightarrow 0}A_\mu=0$. The same prescription works both for the scalar and Dirac
fields. Such potentials can always be found if $\lim_{\eta \rightarrow 0}F_{\mu\nu}=0.$
However,\, one should be aware that our proof only states the identity between amplitudes
in different gauges for which at the infinite future $\lim_{\eta \rightarrow 0}
\Delta A_\mu=0$. One could speculate from here that beside the
electromagnetic field an extra physical information might be stored in the
potential on the conformal boundary of the dS space at $\eta=0$.

Our discussion so far focused on amplitudes in an external classical field. Let us make
a few comments on the question of gauge invariance in the full interacting theory
with $A_\mu$ a quantized field. An immediate observation is that in this case the tree amplitudes
with a single vertex still have the general form ({\ref{auvgen}), with the only difference that the
external potential is replaced by the wave function which corresponds to the photon line. It follows then
with the same argument given in Sec.~3 that these amplitudes are also gauge dependent
quantities.\footnote{Amplitudes of this form involving integrals
identical or very similar to Eq. (\ref{fint}) appear in Refs.~\citen{cruc0,cruc1,nico0,cruc3,
cota0,miha,cruc4,blag1,blag2}. As we mention below, a possible way to guarantee the physical
relevance of these results is that there exists a preferential gauge to perform the calculations.
A natural choice would be the radiation gauge, which is indeed the case in all these papers.}

Another threat to gauge invariance can come of course from the form of the photon
propagator. Given the gauge dependence at the tree level, one can naturally suspect that it
will reappear in higher order calculations, and thus be a generic feature of theory. As we remarked
in Sec.~1, a possible example is provided by the one loop calculations in Refs.~\citen{kahy1, kahy2}.
Let us briefly enumerate some options  in the face of this possibility.

One view would be to\, simply accept that gauge dependence is an unavoidable feature of the
theory in the expanding dS space, as long as one insists on keeping the background fixed. This
could be seen as an unfortunate effect implied by the idealization of an exponential expansion up
to infinite times. As suggested by many authors, it is plausible that in a real
life scenario quantum backreaction effects will slow down  or completely eliminate the expansion, so
that in a more complete theory gauge invariance could be recovered. A similar view was adopted in
the discussion of the unitarity violation in dS space.\cite{nonu}

Another possibility is that the problem is more of a technical nature. For example,
one could suspect that gauge dependence in the amplitudes considered here is related to
the fact that the $out$ modes do not describe real physical states. However, this cannot
be so, which can be seen in the following way. Assuming distinct sets of $in$ and $out$ modes,
the amplitudes for the transition between the initial vacuum and a final state $\alpha$ is
of the form\cite{birr}
\bq
{\cal A}_{0\rightarrow \alpha} = \langle \alpha\, out \vert S \vert 0\, in \rangle
=\sum _\beta \langle \alpha\, out \vert \beta\, in \rangle
\langle \beta\, in \vert S\vert 0\, in \rangle,
\label{sumbog}
\eq
where the scalar products in front of the $S$-matrix elements are defined by the Bogolubov coefficients
which connect the two sets of modes. It is easy to see that in the case examined here, i.e.
first order approximation and $\alpha$ a particle-antiparticle state, only two terms appear
in the sum:\, one which contains the matrix element $\langle \alpha\, in \vert S^{(1)}\vert
\ 0\, in \rangle$ and which represents the amplitudes in Sec. 2, and another one which contains the
vacuum-to-vacuum amplitude $\langle 0\, in \vert S\vert 0\, in \rangle$. The last amplitude comes
multiplied by the factor $\langle \alpha\, out \vert 0\, in \rangle$,\, which due to the
translational invariance of the initial vacuum is proportional to the delta function $\delta^3
({\bf p}+{\bf p}^\prime)$. For an arbitrary external potential $A_\mu(\eta, {\bf x})$ such a
function will $not$ appear in the first term, so that it is impossible
for the gauge variations from the two terms to cancel each other.

One can also contemplate the possibility that gauge dependence is\, a perturbative artifact. It became
clear in recent years that infrared divergences in individual Feynman diagrams in dS space can be
eliminated by resummation techniques (see e.g. Refs.~\citen{riot,arai,yous1}), so that something
similar could happen for the gauge variations of the amplitudes.

Finally,\, a simple way out would be that there exists a preferred gauge which ensures physically
meaningful results. Our restriction (\ref{conpot}) for assuring the gauge invariance of the amplitudes
would then suggest to choose a gauge in which the propagators in conformal coordinates vanish when one
of the points is at future infinity, i.e.
\bq
\lim_{\eta\rightarrow 0}
\Delta_{\mu \nu^\prime}(\eta, \eta^\prime; {\bf x}, {\bf x}^\prime)=0,
\label{limpro}
\eq
and similarly for the primed point. It is essential in this context to recall that the behavior
of the\, dS photon propagators at large spacetime separations can significantly depend on the choice of
gauge.\cite{yous} As shown in the cited paper, among the dS invariant propagators in the
$R_\lambda=\frac{1}{2} (\nabla_\mu A^\mu)^2$ gauges only the propagator in the Landau gauge
$\lambda \rightarrow \infty$ decreases at large spacetime separations. One can also contruct\cite{yous}
other dS invariant propagators whose transverse (physical) part has the same property, but it is not
clear what type of gauge generates them. Unfortunately, one finds\footnote{A collection of formulas which
allow to easily translate the dS invariant propagators\cite{yous} in conformal coordinates can be found in
Refs.~\refcite{kahy1,tsam}.} that even these well behaved propagators do not respect Eq. (\ref{limpro}).
One can also check that the same is true for the various propagators used in
Refs.~\refcite{kahy1}-\refcite{wang2}. For the moment, we do not know whether propagators that
satisfy the property above exist.

As a final suggestion, it would be interesting to try to obtain the dS photon propagator
following the familiar quantization in the radiation gauge in flat space.\cite{bjor} In this case,
one starts with the transversal propagator of the free field, to which one adds the contribution
which accounts for the Coulomb interaction, which after eliminating a pure gauge leads to the Lorentz
invariant propagator in the Feynman gauge. It would be worthwhile to check if going through the same
steps in dS space one can construct a propagator (whether dS invariant or not) which respects
Eq. (\ref{limpro}).

\appendix

\section{}

We obtain here the correspondent of the\, adiabatic residue (\ref{ards}) for
scalar QED. Introducing
\bq
\bar\nu=\sqrt {\frac{m^2}{H^2}-\frac{9}{8}},
\eq
the scalar modes of a definite momentum ${\bf p}$ are given by\cite{birr}
\bq
\varphi_{\bf p}(\eta, {\bf x})=
\frac{\sqrt{\pi/H}} {2(2\pi)^3} \times(-H\eta)^{\frac{3}{2}}\, e^{\frac{\pi \bar \nu}{2}}
H_{-i\bar\nu}^{(1)} (-p\,\eta)\, e^{i{\bf p} \cdot {\bf x}}.
\eq
We suppose this time that $\frac{m}{H}$ is sufficiently large so that $\bar\nu$ is
real. (For $\bar\nu$ imaginary the modes become non-oscillatory for $\eta\rightarrow 0$,
which makes them inappropriate for a description in terms of particle states at late
times. In addition, in this case the amplitudes given below can diverge.\cite{akhm1}) The
amplitudes analogous to Eq. (\ref{genamp}) are
\bq
{\cal A}({\bf p}, {\bf p}^\prime)=-ie \int d^4 x \sqrt {-g }\,g^{\mu\nu}
(f^*_{{\bf p}^\prime}\,i\!\stackrel{\leftrightarrow}{\partial_\mu} f_{\bf p})
A_{\nu},
\label{ampsca}
\eq
with the initial and final wave functions equal to $\varphi_{\bf p}$ or $\varphi_{\bf p^\prime}^*$.
The gauge variations of the amplitudes using a decoupling of the interaction analogous to
Eq. (\ref{vards}) are
\bq
\Delta {\cal A}({\bf p}, {\bf p}^\prime)=ie \int d^4 x \sqrt {-g }\,
h^\prime_\epsilon(t)\, g^{\mu 0}
(f^*_{{\bf p}^\prime}\,i\!\stackrel{\leftrightarrow}{\partial_\mu} f_{\bf p})\,
\Lambda.
\label{varsca}
\eq
Let us fix $f_{{\bf p}^\prime}^*=\varphi^*_{{\bf p}^\prime}$. For precision sake, we will now make
the distinction between $(I)$ creation-annihilation amplitudes, when $f_{\bf p}=\varphi_{\bf p}^*$,\,
and $(II)$ scattering amplitudes, when $f_{\bf p}=\varphi_{\bf p}$. Repeating the construction in
Sec.~4 one finds that the quantities analogous to Eq. (\ref{rds}) for the two types of amplitudes
can be defined as follows:
\bq
R_{dS}^{(I)}(\epsilon)\equiv\frac{\epsilon}{H}
\int_{-H^{-1}}^0 d\eta (-H \eta)^{\,\epsilon/H}
[H^{(2)}_{+i\bar\nu}(-p^\prime\eta)
\stackrel{\leftrightarrow}{\partial_\eta} H^{(2)}_{+i\bar\nu}(-p\,\eta)],
\label{rs2}
\\
R_{dS}^{(II)}(\epsilon)\equiv\frac{\epsilon}{H}
\int_{-H^{-1}}^0 d\eta (-H \eta)^{\,\epsilon/H}
[H^{(2)}_{+i\bar\nu}(-p^\prime\eta)
\stackrel{\leftrightarrow}{\partial_\eta} H^{(1)}_{-i\bar\nu}(-p\,\eta)].
\label{rs1}
\eq
From the experience of the previous calculation, we know that for the result in the limit
$\epsilon\rightarrow 0$ it is sufficient to keep in the power expansion of the Hankel
functions only the terms $\sim \eta^{-1}$. For obtaining these terms it is sufficient
to use
\bq
H^{(1)}_{\nu}(z)\simeq \frac{i}{\sin \pi \nu}
\left\{
\frac{e^{-i\pi \nu}}{\Gamma(1+\nu)}
\left(\frac{z}{2}\right)^\nu
-\frac{1}{\Gamma(1-\nu)}
\left(
\frac{z}{2}\right)^{-\nu}
\right\}, \quad z\rightarrow 0,
\eq
together with $H^{(2)}_{\nu}(z)=H^{(1)}_{\nu}(z)^*$. A calculation similar to that in
Eq. (\ref{ards}) then leads to
\bq
\lim_{\epsilon\rightarrow 0}R_{dS}^{(I)}(\epsilon)
=\frac{2}{i\pi}\frac{e^{-\pi \bar\nu}}{\sinh \pi \bar \nu}
\left\{
\left(\frac{p^\prime}{p}\right)^{-i\bar\nu}-
\left(\frac{p^\prime}{p}\right)^{i\bar\nu}
\right\},
\qquad\,\,\,\,
\label{rds2}
\\
\lim_{\epsilon\rightarrow 0}R_{dS}^{(II)}(\epsilon)
=\frac{2}{i\pi}\frac{1}{\sinh \pi \bar \nu}
\left\{
\left(\frac{p^\prime}{p}\right)^{-i\bar\nu}-
e^{-2\pi \bar\nu} \left(\frac{p^\prime}{p}\right)^{i\bar\nu}
\right\}.
\eq
The first expression is the analogue of Eq. (\ref{ards}) (actually the latter result
is not the full expression due to the simplification in Eq. (\ref{casint})). Notice the
extra factor $e^{-\pi \bar\nu}$ in Eq. (\ref{rds2}), which implies a larger value for the
adiabatic residue $(II)$. This could be translated by saying that the scattering amplitudes
are more sensitive to gauge transformations than the creation-annihilation amplitudes. The
same can be checked to be true for the Dirac field: one finds that for scattering amplitudes
the coefficient $A_H$ in the power series analogous to Eq. (\ref{dive}) contains an
extra factor $\sim e^{+\pi k}$, which leads to the same conclusion.

\end{document}